Reply to Comment on "Solution of the Specific Model of Five-body Problem to Investigate the Effective Alpha-Nucleon Interaction in a Partial-Wave analysis"

E. Ahmadi Pouya[1] and A. A. Rajabi[2]

*Physics Department, Shahrood University of Technology, P. O. Box 3619995161-316, Shahrood, Iran*

The present paper is written in response to the comment of M. R. Hadizadeh *et.al* on our original paper "Solution of the Specific Model of Five-body Problem to Investigate the Effective Alpha-Nucleon Interaction in a Partial-Wave analysis" [Acta Physica Polonica B, 48, 1279 (2017)]. In this paper we present our clarifications and statements on the authors of the comment arguments about of the accuracy of the procedure in solution of simple five- and four-body problems. In this regard our arguments turn out to be very efficient mainly discussed from the author's misunderstanding of the issues discussed in the original paper. In fact, the authors of the comment aims to exaggeratedly show that our paper is completely incorrect, but our following statements prove that the authors of the comment statements are inimical and often inconsequential.

Introduction

In the original paper [1] entitled "Solution of the Specific Model of Five-body Problem to Investigate the Effective Alpha-Nucleon Interaction in a Partial-Wave analysis" in order to study a simple 5- and 4-body bound systems, we have solved the 5-body system in the general case as a bound system within the Yakubovsky approach [2], starting from a typical non-relativistic 5-body Schrodinger equation, Eq. (1.1). Clearly this work is not a simplification of the six-body Yakubovsky equations! (But it is reasonable that by removing the sixth nucleon in the general six-body Yakubovsky equations, the coupled equations leads to 5-body ones). We have solved the coupled integral equations of 5- and 4-body systems to calculate the binding energies of simple specific model of 5-body in the picture of Alpha-N first and then simple 4-body as an inert Alpha particle. Also we have compared our obtained binding energy results with results of other methods with respect to the regarded nucleon-nucleon spin-independent potential models. Finally, we have suggested that some obtained binding energy differences between the 4-body in the case of alpha-particle and the 5-body in the picture of alpha–nucleon system is attractive and of about 13 MeV.

However, as research is generative, every discovery brings about further questions. In the first place, we would like to express our gratitude for the authors of the comment meticulous scrutiny into the article. Secondly, we are comprehensively dedicated to present our reply to the comment in order to compromise any misunderstandings apparent within the comment:

-First, the published paper [3] has been retracted because only the possibility that *part of the published results* were obtained in collaboration with scientist not included in the author list. An erratum to this article is available at https://doi.org/10.1140/epjp/i2017-11724-1.

Therefore, the repetitive authors of the comment arguments about retraction of the paper are completely wrong.

---
[1] E.Ahmady.ph@ut.ac.ir
[2] A.A.Rajabi@shahroodut.ac.ir

As we have discussed and referred in throughout of the published paper [1], again we have addressed and clarified in the theory of the project, introduction, genuine formalism, and the numerical implementation as follows:

At the first place, the main purpose of the published paper is the solution of the five-body bound system as a toy model for five identical particles with considering pairing interactions alike similar methods for bound state of 5-body systems with simple potential models [4-7]. In Sect. 2 the Yakubovsky formalism for 5-body bound system is exactly derived. So, the manner of the formalism in the published paper is mathematically correct and reasonable. Secondly, as we all know and according to the paper "a realistic five-nucleon problem is not allowed for a bound state. However, in order to investigate the effective interactions between the two particles, namely a subsystem as a four-body alpha-particle and an nucleon, we would be considered the five-body problem for the model of effective Alpha–Nucleon structure as a bound system. Also, an objection to the use of simple phenomenological potentials for Alpha-Nucleon scattering arises from the fact that these potentials allow a bound state for the 5-body system which is forbidden by the exclusion principle. Therefore, in order to calculate the effective Alpha-Nucleon model in the special specific structure of the five-body model system, we study the Yakubovsky scheme, extending the applications to systems with $A = 5$ (line 4 in the third paragraph to the Introduction). Also in the entire paper we have not mentioned that our system is $^5$He or $^5$Li, because the spin/isospin effects are not taken into accounts. Therefore, there is no difference between proton and neutron interactions in the simple 5-body system. Also we have not mentioned in the paper that the simple specific Alpha-Nucleon system is a Halo structure system.

1. In response to the comments on the Title of the paper:

(a) We formulate the general 5-body problem within the Yakubovsky approach [2]. According to the 4 coupled equations in terms of 4 independent components we find that just two first components (configurations) are related to 4N-N, namely simple structure of Alpha-Nucleon specific model. So we have chosen the first two relevant configurations, and the other configurations will not be taken into accounts in the study of specific 4N-N structure. Moreover, according to Fig. 1, the effective interaction of alpha-particle is governor and concealed in the remained components (see first two configurations in Fig. 1 and compare them with Figs. 1 and 2 in Ref. [8]). A good reason of the accuracy of formalism is removing the interactions of the fifth nucleon in the specific 4N-N coupled equations, because the five-body problem leads to a typical four-body problem [8]. Notice, the simple 5-body system in the case of Alpha-Nucleon structure has a 4-body sub-system as a simple alpha particle. So, according to the results, some obtained binding energy differences between the 4-body as alpha-particle and the 5-body as alpha–nucleon system we suggested that an effective interaction of Alpha and Nucleon is attractive and of about 13 MeV.

(b) As we have discussed in Sect. 4 and Appendix C, it is obvious that the integral kernels are evaluated within the Partial-Wave approach, not 3D, because Eq. (B.5) and (B.6) and similarly (4.1) and (4.2) are scalars not vectors! However, in order to reduce the high-dimension of problem we can choose suitable coordinate systems. In 3D approach the basis states are vectors, but the momentum basis states in the paper are scalars.

2. In response to the comments on the Abstract:

(a) The authors of the comment about "The Yakubovsky formalism for a 5-nucleon system leads to a set of 5 coupled equations" according to ref. [9] is completely wrong and the fig. 1 in ref. [9] is completely compatible with fig. 1 in our paper [1]. The five independent structures not mean just five coupled equations, because the last component is referred to two linear combinations of two different structures and our obtained structures of the 5-body system in the picture of Jacobi coordinates are completely correct. Also how the coupled equations of 5-body most be equal with coupled equation of 6-body ones [10]. Does the statements of the authors of the comment is correct?! Of course not.

(b) We have absolutely not mentioned that the two first structures are halo! These structures describe just the simple Alpha-Nucleon (4N-N) structures, not a halo namely $^5$He. Also in the first step to the calculations of the 5-body system, we considered the simple specific toy model of 5-body problem.

(c) We suggested that the binding energy differences between the 4-body in the model of Alpha and 5-body in the model of Alpha-Nucleon structures can be described as a simple effective interaction between an inert alpha-particle and a nucleon that is attractive and of about 13 MeV.

Our obtained structures [1]                other obtained structures [9]

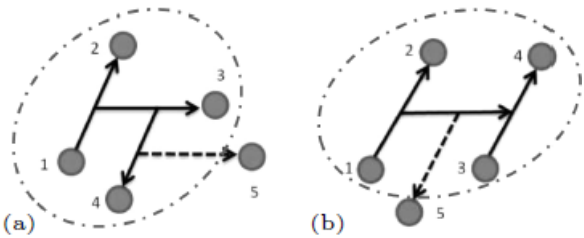 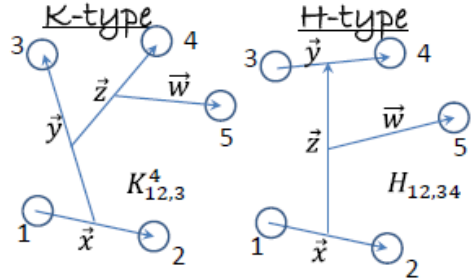

Fig. 1. The two types of Alpha-Nucleon models [(a) and (b) / K-type and H-type] of the Yakubovsky components in the picture of 5-body Jacobi coordinates. Obviously, there is no difference between left and right structures. Also, in the bag, the 4-body subsystem plays a simple alpha model [8]. It is clear that the 4-body as an alpha particle have attractive interaction with a nucleon.

(d) The authors of the comment argue that our results with spin-independent potential model have been compared with obtained results from other methods with spin-dependent potentials! The authors of the comment don't see Sect. 5 and other method references of the paper! In sect. 5 of the paper we have represented the applied spin-independent potential models and we have presented and compared our obtained results with results of other methods, with respect to the regarded applied spin-independent interactions. Also the Tables of the results are separated in terms of the kind of potentials. All listed results in the published paper, are obtained from applying spin-independent potentials.

3. In response to the comments on the Introduction:

(a) According to the first paragraph in introduction "the investigation of light nuclei and the study of the identity of the governing effective interactions, in addition to the specific properties of the bound and scattering states, are very interesting and relevant topics in nuclear few-body systems, as well as the atomic community. The main interest in the few-body problems is finding an accurate solution for the systems, as well as looking for unknown interactions governing these systems. To this regard, the investigation of few-nucleon bound systems interacting via simple and realistic interactions has been always in the center of interest. The description of light nuclei and the effective Alpha-Nucleon interactions especially require well-established methods to solve the non-relativistic Schrödinger equation, in addition to the description of the relevant models of such interactions." Are completely relevant, and the motivation of the paper is the study of the 5-body problem in the case of specific 4N-N model can be described as a simple Alpha-Nucleon system.

(b) In this case, the technical performance in Partial-wave decomposition is implemented and 2 different set Jacobi momenta as well as a necessary multi-dimensional interpolation scheme are given, like modified cubic Hermit splines.

(c) It means that we have investigated the convergence of the eigenvalue of the Yakubovsky kernel with respect to the number of grid points and calculate the expectation value of the Hamiltonian operator, which is efficient with respect to the number of components and well-suited for a numerical implementation.

(d) We have found it natural to use the standard PW representation, essentially because we are dealing with scalar variables. The partial-wave representation of the basis states and evaluation of integral kernels are given in Appendix B and C, explicitly. Also we have considered the nucleons as fermionic spinless particles, in L=1 (The Pauli principle is taken into accounts $\varphi = -P_{ij}\, \varphi$), therefore before evaluation of scalar equations (partial-wave method), in order to reduce the high-dimension of the variables we have selected a special coordinate systems and L=1 is considered to this selection.

4. In response to the comments on the Formalism:

(a) It is well-known in order to the solution of 5-body problem within the Yakubovsky scheme we have started with the 5-body Schrodinger equations, and there is not any simplification to the formalism. Moreover, in the first step to the formalism, we have cited the "*The Quantum Mechanical Few-Body Problem*, Springer published book by Glöckle [11] that in which some few-body systems A<=4 within the Faddeev-Yakubovsky methods have been described. Also, we have referred the Faddeev scheme in Eq. (2.2) and the Lippmann-Schwinger operator forms in Eq. (2.4) due to the formalism.

(b) The two coupled equations of the specific 5-body problem, namely Eqs. (3.1) and (3.2), are related to the very approximate effective two-body configuration of Alpha–Nucleon model, Fig 1. Therefore the other components are not taken into accounts. So, the simplification of the general 6-body system is not transferred to our specific 5-body problem. But, according to Fig. 1, the effective 4-body subsystem as an alpha-particle is governor and concealed in the two remained components (similarly two coupled equations). By removing the fifth nucleon interactions in the remaining two coupled equations ($P_{45}$ equal to be zero) the specific model of 5-body problem leads to a typical 4-body problem [8]. To this aim, against the authors of the comment arguments, the 5-body Jacobi configurations of ref. [9] and our paper [1] is represented, respectively, as follows:

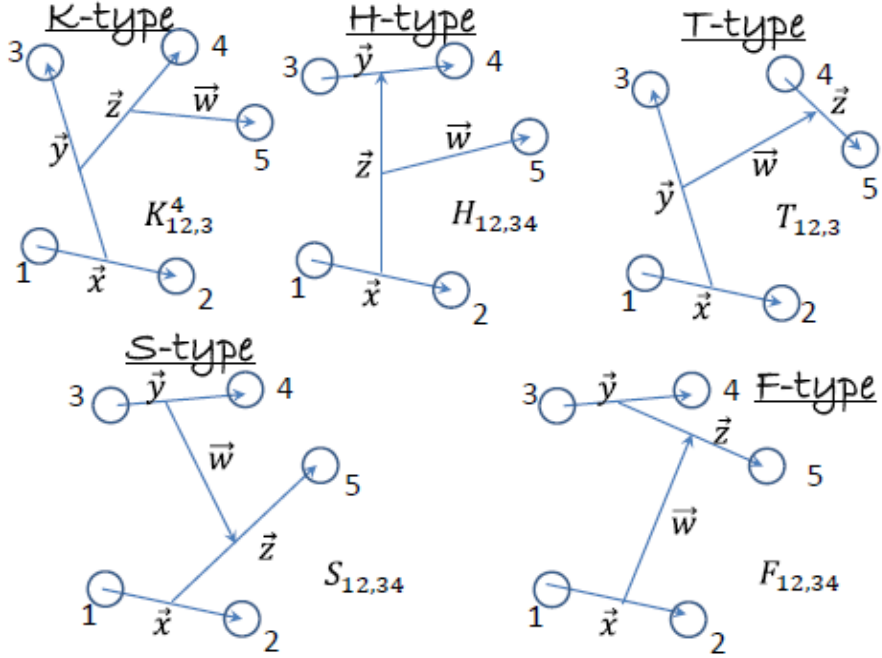

FIG. 1: (Color online) 5-particle Jacobi coordinate sets used to describe FY components, denoted in this work as $K, T, H, S, F$.

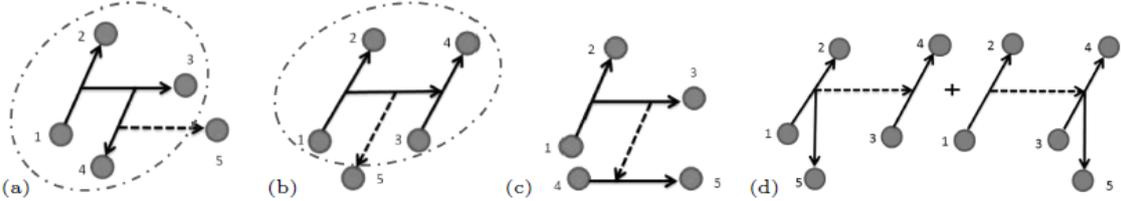

Fig. 1. Diagramatic configuration of the four independent components of 5-body system in Jacobi coordinates. (a), (b), (c) and (d) are respectively the configurations of the components $\varphi_{12;123}^{1234}$, $\varphi_{12;12+34}^{1234}$, $\varphi_{12,123}^{123+45}$ and ($\varphi_{12,12+34}^{125+34} + \varphi_{12,12+34}^{12+345}$). In the bag, the alpha–core plays as a 4-body subsystem.

Fig. 2. Schematically comparison of our obtained 5-body configurations [1] with other 5-body configurations [9].

As we all know the two types of configuration are clearly compatible. Also against the authors of the comment arguments in ref. [9] the coupled 5-body equations is not mentioned and derived, See ref. [9] for details. Therefore, our configurations and 4 coupled equations from derivation of Schrodinger equation, mathematically, are completely correct and the authors of the comment arguments about the results of formalism similar to previous repetitive arguments are not valid.

(c) In the published paper the major of the nucleons are considered as fermionic spinless-particle. See paragraph after Eqs. (3.1) and (3.2), "It is well-known that such a nuclear system should be treated in the fermionic approaches, i.e. the five-body total wave function follows $\varphi = -P_{ij}\,\varphi$, and the Pauli principle is taken into accounts, even for spinless particles". Therefore, the total wave-function of the specific 5-body is obtained antisymmetric function, and the specific 5-body coupled equations, namely Eqs. (3.1) and (3.2), are correct, against the authors of the comment arguments.

(d) Clearly the spin effects are different from the angular momentum degrees of freedom. Therefore, we switched off the spin effects in the first step to the calculations, but the angular momentum effects are considered to the selecting coordinate systems. After Eq. (B.6) in appendix B" we describe dependent on angular grid points by choosing the relevant coordinate systems, because total angular momentums of the 5-body system are restricted in L = 1 state, and the Pauli principle will be taken into accounts (see Sect. 4 in the paper [1])."

(e) The authors of the comments even don't see that in addition to the calculations of the 5-body problem, in order to investigate the effective Alpha-nucleon interaction, we have calculated the 4-body bound state as the simple Alpha particle model system, Eqs.(3.3) and (3.4). Also we have considered the T-operators with compatible for 5-body and 4-body equations, namely for 5-body the transition operators are $T^{123+4+5}$ and $T^{12+34+5}$ and for 4-body are $T^{123+4}$ and $T^{12+34}$ and we have notated that in the sub-cluster notation the single particle, namely fifth particle, will no longer be displayed. Therefore the authors of the comment think that the T-operators for 5- and 4-body problems are equal in action.

5. In response to the comments on the numerical implementation

(a) After selecting the suitable components, we have described the corresponding Jacobi momentum in standard manner, explicitly. Next we have represented the corresponding momentum basis state in terms of the K- and H-set Jacobi momenta, respectively. It is clear that the basis states Eqs. (B.5) and (B.6) are obviously scalars not vectors! Against the authors of the comment arguments, because we have represented and evaluated the projected coupled equations within the Partial-wave method (namely scalar basis states not 3D basis). Selecting the coordinate system not means that we use 3D method. In appendix C the integral kernels are clearly evaluated in Partial-wave manner. But just for reduce the high dimension of the problem we have projected the momentum vectors in special coordinate system, but the basis states was done in scalar state. How the authors of the comment reach by this standard manner, the calculations have not done! Meanwhile we have surprised that their arguments especially this part are not scientific and are emotional comments.

(b) The authors of the comment believe that the 7-10 iterations is enough and we have calculated that the 10-20 iterations is sufficient, there is no conflict about the issue, because we have calculated the problem with 10 iteration numbers. In order to accuracy of the calculations, in sect. 5.2, we have implemented the numerical stability of our algorithm and our representation of the 5-body Yakubovsky components in Partial-wave analysis. We have specially investigated the convergence of the eigenvalue of the Yakubovsky kernel with respect to the number of grid points for Jacobi momenta, azimuthal and spherical angle variables.

(c) In ref. [8] the similar details of the typical Pad'e approximation for 4-body Transition-operators are represented, explicitly. Therefore in this paper is not reasonable the definition of a typical iteration Pad'e method, for avoiding the long formulation. Also we have cited the relevant paper [8].

6. In response to the comments on the numerical results

   (a) The applied potential in our method and other methods are spin-independent models, how authors of the comment have argued that other methods have been applied with some spin-dependent potential. This calculation and other calculation have been applied with spin-independent interaction, namely Volkov and Malfliet-Tjon V, so there is no difference between our obtained results and other results in terms of spin effects. Also as we have mentioned in the published paper, the full contributions of four nucleons in 4-body as inert subsystem completely are dedicated in the remaining components, and the binding of the fifth nucleon is adaptable with full contributions. Therefore, our obtained results are in fair compatible with results of other methods, with respect to the regarded spin-less interactions.

   (b) In response to the repetitive comments about the 13 MeV, in conclusion we suggested that the binding energy differences between the 4-body in the model of Alpha and the 5-body in the model of Alpha-Nucleon structures can be described as a simple effective interaction between an inert alpha-particle and a nucleon that is attractive and of about 13 MeV.
       i.    According to Fig. 1 the binding energy differences between 4-body as alpha-particle and 5-body as Alpha-nucleon system, indicates the interaction of fifth nucleon with a compact 4-body subsystem in such a specific model, is attractive and of about 13 MeV, that we described the Alpha-nucleon interaction.
       i.    We have just expected that for considering the spin-dependent potential models, this interaction will not be changed, because the both binding energies will be improved equally." It is worthwhile to mention that by including the spin effects in the implementation of the four-body system in the model of alpha-particle and five-body system in the specific model of effective Alpha–Nucleon structure, both binding energy results will be

almost equally improved, so correspondingly, the results of the effective Alpha-Nucleon interaction will remain almost unchanged when the spin-dependent interactions are used." (Second paragraph in conclusions).

(c) We avoided the total wave function presentation of the specific 5-body model in the published paper. But in response to the comment, we represent the explicit form of the total 5-body wave-function in the case of Alpha-Nucleon specific model. To this regard we have used Eqs. (2.1), (2.5), (A.19), (A.27) of the paper, together with adequate permutation operators and leads:

$$\begin{aligned}\Phi_{\alpha-N} = \quad & [1 - P_{23} - P_{24} - P_{13} - P_{14} + P_{13}P_{24}]\big[(1 - P_{34})\varphi^{1234}_{12;123} + \varphi^{1234}_{12;12+34}\big] \\
& -[1 - P_{23} - P_{24} - P_{13} - P_{14} + P_{13}P_{24}] \\
& \times \big[((1 - P_{34})(P_{45} + P_{35}))\varphi^{1234}_{12;123} + (P_{45} + P_{35})\varphi^{1234}_{12;12+34}\big] \\
& -[P_{25} + P_{15} - P_{13}P_{25} - P_{14}P_{25}] \\
& \times \big[(1 - P_{34})\varphi^{1234}_{12;123} + \varphi^{1234}_{12;12+34} - ((1 - P_{34})(P_{45} + P_{35}))\varphi^{1234}_{12;123} \\
& -(P_{45} + P_{35})\varphi^{1234}_{12;12+34}\big].\end{aligned} \quad (1)$$

(d) As you know in ref. [12] the spin and all angular momenta degrees of freedom have been considered for 3-nucleon bound state calculation, but in the calculations of the 4-body system and 5-body problem [1] with 4 significant digits, dedicating 20 grid points are enough. However, a better agreement could be reached if we considered a larger number of grid points in our calculations.

**Conclusions**

According to the main part of the paper to the formalism, we have formulated and mathematically derived the general 5-body problem within the Yakubovsky approach that is completely genius and can be applied for the study of each 5-body problem in various quantum mechanical fields, such as simple five-nucleon problems, constituent quark-model (penta-quarks) and atomic bound models (pentamers). Also, the explicit evaluation of the 4- and 5-body Yakubovsky coupled integral equations are implemented correctly in a Partial-wave analysis, and the eigenvalue equations are solved for specific Alpha-nucleon (5-body) and alpha (4-body) model systems. In addition, the stability of our algorithm has been achieved with the calculation of the eigenvalue of Yakubovsky kernel, where a four number of grid points for Jacobi momenta and angle variables have been used. We have also calculated the expectation value of the 5-body Hamiltonian operator. It is worth to mention that all steps of the published paper are the first step toward the calculations of 5-body problems and every discovery brings about further questions and criticism comments.